\documentclass{lmcs}
\pdfoutput=1

\usepackage{lastpage}
\lmcsdoi{16}{4}{10}
\lmcsheading{}{\pageref{LastPage}}{}{}%
{Jul.~18,~2016}{Nov.~12,~2020}{}

\usepackage[utf8]{inputenc}

\author{Mathys Rennela}
\address{Centrum Wiskunde \& Informatica, Amsterdam}
\email{mathys.rennela@gmail.com}


\title[Convexity and Order in PFPC]{Convexity and Order in Probabilistic Call-by-Name FPC}

\usepackage{amsthm,amsmath,amssymb}
\usepackage{mathtools}
\usepackage{bm}
\usepackage{cite}
\usepackage[all,cmtip]{xy}
\usepackage{appendix}

\usepackage{stmaryrd}
\usepackage{comment}
\usepackage{subfig}
\usepackage{microtype} 

\usepackage{float}
\floatstyle{boxed}
\restylefloat{figure}

\newcommand{\hide}[1]{}

\newcommand{\lub}{\bigvee}
\newcommand{\abs}[1]{\left| #1 \right|}

\newcommand{\nattrans}{\Rightarrow}

\newcommand{\myvec}{\overrightarrow}

\newcommand{\denot}[1]{[\![#1]\!]}

\newcommand{\QQ}{\mathbb{Q}}

\newcommand{\cat}[1]{\mathbf{#1}}

\newcommand{\opp}[1]{#1^\mathbf{op}}
\newcommand{\yoneda}{\textbf{y}}

\newcommand{\defeq}{\stackrel{\text{def}}=}

\newcommand{\kl}{\text{Kl}}

\newcommand{\Fd}{\mathbf{Fd}}

\newcommand{\CFDCPU}{\Fd\mathbf{C}\CStar_{\mathrm{CPU}}}

\newcommand{\Conv}{\cat{Conv}}
\newcommand{\dConv}{\cat{dConv}}
\newcommand{\Set}{\cat{Set}}

\newcommand{\op}{\textbf}
\newcommand{\opcase}[3]{\op{case}(#1, x \cdot #2, y \cdot #3)} 

\newcommand{\unit}{[0,1]}
\newcommand{\R}{\mathbb{R}}
\newcommand{\N}{\mathbb{N}}

\newcommand{\Dcpo}{\mathbf{Dcpo}}
\newcommand{\DcpoS}{\Dcpo_{\perp!}}

\newcommand{\CStar}{\mathbf{C^*\-Alg}}

\newcommand{\eppair}[2]{\left\langle#1,#2\right\rangle}
\newcommand{\id}{\operatorname{id}}
\newcommand{\zero}{\mathbf{0}}

\newcommand{\DM}{\mathcal{D}}

\newcommand{\monad} T

\newcommand{\dSConv}{\cat{d}\SConv}
\newcommand{\SConv}{{\Conv_{\leq 1}}} 
\newcommand{\KS}{\cat{KS}}
\newcommand{\Lawv}[1]{\mathbb{L}_{#1}}
\newcommand{\LawvConv}{\Lawv{}}
\newcommand{\LawvSConv}{\Lawv{\leq 1}} 
\newcommand{\LawvOptConv}{\Lawv{(\leq 1)}} 
\newcommand{\SDM}{\mathcal D_{\leq 1}} 
\newcommand{\DMopt}{\mathcal D_{(\leq 1)}} 
\newcommand{\DMinf}{\DM^\infty}
\newcommand{\SDMinf}{\SDM^\infty}
\newcommand{\DMinfopt}{\DMopt^\infty}

\newcommand{\nat}{{\op{nat}}}

\newcommand{\num}[1]{\underline{#1}}
\newcommand{\xprobred}[1]{\xrightarrow{#1}}
\newcommand{\probred}{\xprobred{\kappa}}
\newcommand{\detred}{\to_d}
\newcommand{\ifop}{\op{if}}
\newcommand{\fixop}{\op{fix}}
\newcommand{\Prob}{\op{Prob}}

\newcommand{\arrow}{\multimap}
\newcommand{\homDay}[2]{[#1,#2]_\text{Day}}
\newcommand{\Day}{\otimes_\text{Day}}

\newcommand{\tif}{\text{if}}
\newcommand{\tnat}{\text{nat}}
\newcommand{\tsucc}{\text{succ}}
\newcommand{\tcoin}{\text{coin}}
\newcommand{\pcfnum}[1]{\underline{#1}}
\newcommand{\tfix}{\text{fix}}

\usepackage{prooftree}

\newcommand{\kindoftheorem}{Theorem}
\newtheorem*{theoremcustom}{\kindoftheorem}

\renewcommand{\DM}{\mathcal{D}}

\usepackage{listings}
\lstdefinestyle{customc}{
  belowcaptionskip=1\baselineskip,
  breaklines=true,
  frame=L,
  xleftmargin=\parindent,
  language=Python,
  showstringspaces=false,
  basicstyle=\small\ttfamily,
  tabsize=2,
  breakatwhitespace=false,
  breaklines=true,
  captionpos=b,
  frame=none,
  keepspaces=true,
  morekeywords={box,case,of,in1,in2,output,gate,unbox,run,lift},           
}
\newcommand{\symcat}[1]{\ensuremath{\vert #1 \vert}}
\renewcommand{\arrow}{\to}

\begin{document}
\maketitle

\begin{abstract}
Kegelspitzen are mathematical structures coined by Keimel and Plotkin, in order to encompass the structure of a convex set and the structure of a dcpo. In this paper, we ask ourselves what are Kegelspitzen the model of. We adopt a categorical viewpoint and show that Kegelspitzen model stochastic matrices onto a category of domains. Consequently, Kegelspitzen form a denotational model of pPCF, an abstract functional programming language for probabilistic computing. We conclude the present work with a discussion of the interpretation of (probabilistic) recursive types, which are types for entities which might contain other entities of the same type, such as lists and trees.
\end{abstract}

\bigskip
The interplay between convexity and order in the semantics of probabilistic programs has been a highly-coveted field of research since the first research programs~\cite{jones-thesis,jones-plotkin-powerdomain} on the semantics of probabilistic computing, a programming language paradigm which allows probabilistic branching of programs and also updating of distributions.

Starting from an intuitive and minimalistic programming language perspective on Keimel \& Plotkin's approach to probabilistic computations~\cite{keimel-plotkin-kegelspitzen}, the present work provides a new take on the mathematical characterization of probabilistic programs and brings an important building block to the study of the interactions between the concepts of convexity and order within the theory of probabilistic computing, namely by defining Kegelspitzen as mathematical structures which combine convex sets with dcpos.

We introduce Kegelspitzen as pointed dcpos with a compatible convex structure which carries a clear probabilistic interpretation (see Section~\ref{sec:B1:Kegelspitzen}). We pursue in Section~\ref{sec:B1:categorical-account-Kegelspitzen} with a categorical study of Kegelspitzen, which was absent from Keimel \& Plotkin's original work~\cite{keimel-plotkin-kegelspitzen}.

Now, recall that (sub)convex sets are sets equipped with a (sub)convex structure. After defining the Lawvere theory $\LawvConv$ of convex sets and the Lawvere theory $\LawvSConv$ of subconvex sets, and establishing that those categories have all finite products (see Lemma~\ref{lem:B1:LawvConv-finite-products}), we show the following theorem. 

\renewcommand{\kindoftheorem}{Theorem~\ref{th:KS-LawvSConv-DcpoS}}
\begin{theoremcustom}[paraphrased]
The category of Kegelspitzen and affine Scott-continuous maps, i.e.~Scott-continuous maps which preserve the convex structures, is equivalent to the order-enriched category of models (i.e.~finite product-preserving order-enriched functors) of the Lawvere theory of subconvex sets into the category of pointed dcpos and strict Scott-continuous maps.
\end{theoremcustom}

In a second step, we show that the category of Kegelspitzen and affine Scott-continuous maps is monoidal closed (see Proposition~\ref{prop:B1:KS-mon-closed}), when equipped with the smash product $\otimes_\perp$~\cite{amadio-curien,abramsky-jung}, i.e.~the quotient of the cartesian product $X \times Y$ (of two pointed dcpos $X$ and $Y$) by the relation generated by the relation $\sim$ such that $(x,\perp) \sim (\perp,y) \sim (\perp,\perp)$ for $x \in X$ and $y \in Y$. Moreover, we show that the category of Kegelspitzen and Scott-continuous maps is cartesian closed (see Proposition~\ref{prop:B1:KS-cart-closed}).

Then in Section~\ref{sec:B1:interpreting-pPCF}, we use the cartesian closed structure of the category of Kegelspitzen and Scott-continuous maps to interpret a probabilistic extension called Probabilistic PCF (or shortly, pPCF) of the language PCF~\cite{plotkin-PCF}. In short, we extend PCF with terms $\tcoin(\kappa)$ (where $\kappa \in \unit \cap \QQ$ is a probability) which reduce to the numeral $\num{0}$ with probability $\kappa$ and the numeral $\num{1}$ with probability $1-\kappa$. Therefore, pPCF's transition system is probabilistic: reductions are weighted by probabilities, and deterministic reductions are weighted by the probability $1$.

We proceed to interpret types as Kegelspitzen and terms as Scott-continuous maps. In particular, the type $\nat$ is denoted by the Kegelspitze of sub-distributions on the natural numbers:
\[
\SDMinf(\N) \defeq \left\{\varphi:\N \to \unit ~\middle|~ \sum_{n \in \N} \varphi(n)\leq 1\right\}
\]
We obtain the following soundness property.

\renewcommand{\kindoftheorem}{Proposition~\ref{prop:B1:soundness}}
\begin{theoremcustom}[paraphrased]
The denotation under a context $\Gamma$ of a term $M$ (which isn't a value) is the sum of the denotations under the context $\Gamma$ of the terms that $M$ reduces to.
\end{theoremcustom}

This mathematical observation leads us to the following adequacy result.

\renewcommand{\kindoftheorem}{Theorem~\ref{thm:B1:adequacy}}
\begin{theoremcustom}[paraphrased]
The denotation of a closed term $M$ of type $\nat$ maps every natural number $n$ to the probability that $M$ reduces to the number $\num{n}$ in pPCF's leftmost outermost strategy.
\end{theoremcustom}

We conclude the present work with a proof that the category of Kegelspitzen and affine Scott-continuous maps is algebraically compact for locally continuous endofunctors (see Corollary~\ref{cor:B1:KS-alg-compt}), and as such a model of the language FPC, an extension of PCF with recursive types~\cite{fiore-FPC}: this settles Kegelspitzen as an adequate categorical setting for denoting recursive types.

The semantics of a probabilistic extension of PCF has notably been studied in a similar setting in~\cite{ppcf-angelic}. It is also worth mentioning that previous work proved that probabilistic coherence spaces constitute a fully abstract model of pPCF (see e.g.~\cite{danos-ehrhard-PPCF,computational-meaning-PCS,ehrhard-tasson-pagani-PCS-FA-pPCF,ehrhard-tasson-pCBPV}). Moreover, probabilistic coherence spaces give an interpretation of recursive types based on the relational model\footnote{Recall that in the relational model of linear logic, all linear logic connectives are Scott continuous functions on the class of sets ordered by inclusion.} of linear logic, i.e.~based on the category {$\cat{Rel}$} of sets and relations (see e.g.~\cite{ehrhard-tasson-pCBPV}).

Kegelspitzen offer an interesting categorical semantics within the scope of probabilistic computing, especially as a step towards the study of the semantics for a higher-order quantum programming language with recursive types but also as a subset of the probabilistic fragment of a categorical model of a language for quantum circuits based on C*-algebras (see~\cite{rennela-staton-mfps17}). Indeed,
the category $\CFDCPU$ of finite-dimensional commutative C*-algebras and completely positive unital maps between them is equivalent to the Lawvere theory of convex sets~\cite[Prop.~4.3]{furber-jacobs-calco13}.


\section{An introduction to the theory of Kegelspitzen}%
\label{sec:B1:Kegelspitzen}

In this section, we give a concise introduction to Kegelspitzen, introduced by Keimel \& Plotkin~\cite{keimel-plotkin-kegelspitzen} as pointed dcpos with a compatible convex structure. The word \textit{Kegelspitze} (plural \textit{Kegelspitzen}) is the german term for ``cone tip''.

\medskip
But first, let us recall the formal definition of a convex set.

\begin{defi}\label{def:convset}
A \emph{convex set} (resp.~\emph{subconvex set}) is a set $X$ together with an $m$-ary function
${(\myvec r)}_X:X^m\to X$ for each vector $\myvec r=(r_1\dots r_m)$ of non-negative real numbers
with $\sum_i r_i=1$ (resp.~$\sum_i r_i\leq 1$),
such that
for each $m\times n$ matrix ${(s_{i,j})}_{i,j}$ of non-negative real numbers such that
$\sum_j s_{i,j}=1$, we have
$
\sum_i r_i.(\sum_j (s_{i,j}.x_j))
=
\sum_j ((\sum_i (r_i. s_{i,j})).x_j)$.

A \emph{homomorphism} of (sub)convex sets is a function that preserves the algebraic structure. Homomorphisms are often called \emph{affine maps}. We write $\Conv$ (resp.~$\SConv$) for the category of convex sets (resp.~subconvex sets) and affine maps between them. 
\end{defi}

A \emph{convex dcpo} is a convex set
equipped with a dcpo structure such that the functions that constitute its convex structure are Scott-continuous. A simple example of a convex dcpo is the unit interval $\unit$ of the reals. We will consider the category {$\dConv$} of convex dcpos and affine Scott-continuous maps, i.e.~Scott-continuous functions which preserve the algebraic structure. For two convex dcpos $D_1$ and $D_2$, the homset $\dConv(D_1,D_2)$ can be seen as a dcpo (and is considered as such in this chapter) or as a convex set.

\medskip
A \emph{pointed convex dcpo} (or \emph{subconvex dcpo}) is a convex set and a dcpo with a least element that is a zero element for the convex structure. We will consider the category {$\dSConv$} of pointed convex dcpos and affine strict Scott-continuous maps.

\medskip
A \emph{Kegelspitze} is a pointed convex dcpo $X$ with a convex structure such that the scalar multiplication $\cdot: \unit \times X \to X$, defined by $\lambda \cdot x = x\  \oplus_\lambda \perp$, is Scott-continuous in both arguments. When the unit interval $\unit$ carries the Scott topology, the requirement is that the scalar multiplication is continuous in the product topology of its domain. We will refer to this assumption as the ``Kegelspitzen condition''. The interested reader can consult~\cite{keimel-plotkin-kegelspitzen} for more details.

\medskip
Alternatively, one can define a Kegelspitze as a pointed convex dcpo $X$ with the following properties:
\begin{itemize}
 \item the function $f:\unit \times X^2 \to X$ defined by $f(\lambda,(x,y))=x \oplus_\lambda y$, where $\unit$ is endowed with the usual Hausdorff topology, is continuous in both arguments;
  \item for every natural number $n$, the function $\theta_{n,X}:\SDMinf(n) \times X^n \to X$ defined by
  \[({(\lambda_i)}_{i \leq n},{(x_i)}_{i \leq n}) \mapsto \sum_i \lambda_i \cdot x_i\]
   (where $\SDMinf(n)\cong\{(q_1,\ldots,q_n) \in \unit^n \mid \sum_{i=1}^n q_i \leq 1\}$ carries the Scott topology) is continuous in both arguments
\end{itemize}

\noindent
A \emph{homomorphism of Kegelspitzen} is an affine strict Scott-continuous map of Kegelspitzen. Such homomorphisms are called \emph{affine Scott-continuous maps}. Then, the category {$\KS$} is the category of Kegelspitzen and affine Scott-continuous maps between them. For an historical account of the different notions of Kegelspitzen, see~\cite[Remark 2.28]{keimel-plotkin-kegelspitzen}.


\medskip
Since we intend to use Kegelspitzen as a categorical model for higher-order probabilistic computation, it seems natural to check whether it is a monoidal closed category suitable for the interpretation of recursive types. A step towards this goal requires to give a categorical account of Kegelspitzen, as models of the Lawvere theory of subconvex sets in the category of pointed dcpos and strict Scott-continuous maps.

\section{A categorical account of convexity and order}%
\label{sec:B1:categorical-account-Kegelspitzen}

In this section, we will formally justify the definition of Kegelspitzen by proving that they are models of the order-enriched Lawvere theory of subconvex sets in the category $\DcpoS$ of pointed dcpos and strict Scott-continuous maps. But first, let us recall the preliminary notions involved in our categorical construction of Kegelspitzen.

\begin{defiC}[\cite{jacobs-coalgebra-book}]%
\label{def:B1:distribution-monads}
The monad $\DMinf$ (resp.~the monad $\SDMinf$) is the \emph{infinitary (sub)probabilistic discrete distribution monad} on the category $\cat{Set}$. It is defined as follows on sets: 
\[
\DMinf(X)=\left\{\varphi:X \to \unit ~\middle|~ \sum_x \varphi(x)=1\right\}
\]
\[
\SDMinf(X)=\left\{\varphi:X \to \unit ~\middle|~ \sum_x \varphi(x)\leq 1\right\}
\]
In particular, when $X$ is a finite set of cardinality $n \in \N$, identified with the $n$-element set noted $n$:
\[
\DMinf(n)=\left\{{(x_k)}_{1\leq k \leq n} \in \unit^n ~\middle|~ \sum_k x_k=1\right\}
\]
\[
\SDMinf(n)=\left\{{(x_k)}_{1\leq k \leq n} \in \unit^n ~\middle|~ \sum_k x_k \leq 1\right\}
\]
For every function $f: X \to Y$, the function $\DMinfopt(f): \DMinfopt(X) \to \DMinfopt(Y)$ is defined by:
\[
 \varphi \mapsto \left(y \mapsto \sum_{x \in f^{-1}(y)} \varphi(x) = \sum_{x: f(x)=y} \varphi(x) \right)
\]
The unit $\eta: \text{Id}_X \nattrans \DMinfopt$ and the multiplication $\mu: \DMinfopt\DMinfopt \nattrans \DMinfopt$ are given for every set $X$ by the following:
\begin{align*}
\eta_X: X &\to \DMinfopt X
&\mu_X : \DMinfopt \DMinfopt X &\to \DMinfopt X\\
x &\mapsto \delta_x
&\Phi &\mapsto \left(x \mapsto \sum_{\varphi \in \DMinfopt X} \Phi(\varphi) \cdot \varphi(x)\right)
\end{align*}
where $\delta_x$ is the Dirac notation for $x \in X$, i.e.~for every $y \in X$, $\delta_x(y)=1$ if $x=y$ and $\delta_x(y)=0$ if $x \neq y$.
\end{defiC}

Recall that a \textit{Lawvere theory} is
a small category {$\mathbb{T}$} with (finite) products such that every object is identified with a natural number $n \in \N$
and that a \textit{model} of a Lawvere theory $\mathbb{T}$ is a product-preserving functor $\mathbb{T} \to \cat{Set}$~\cite{lawvere}. More generally, a model of a Lawvere theory $\mathbb{T}$ into a monoidal category $\cat V$ is a tensor-preserving functor $\mathbb{T} \to \cat{V}$.

\medskip
In what follows, we want to construct the categories $\LawvConv$ and $\LawvSConv$ to be the Lawvere theories of the equational theories of convex sets and subconvex sets respectively.
We define $\LawvConv$ (resp.~$\LawvSConv$) as the opposite category of free $\DMinf$-algebras (resp.~free $\SDMinf$-algebras) on finitely many generators. The category $\LawvConv$ (resp.~$\LawvSConv$) is the category with natural numbers as objects together with arrows $n \to m$ seen as probabilistic transition matrices $m \to \DMinf(n)$ (resp.~sub-probabilistic transition matrices $m \to \SDMinf(n)$), i.e.~as stochastic matrices of size $m \times n$, i.e.~$m \times n$ matrices with positive entries such that each column sums up to $1$ (resp.~sums up to a value below or equal to $1$). In the language of monads, this means that $\LawvConv$ (resp.~$\LawvSConv$) is the category $\opp{\kl_\N(\DMinf)}$ (resp.~$\opp{\kl_\N(\SDMinf)}$), i.e.\ the opposite category of the Kleisli category of the monad $\DMinf$ (resp.~$\SDMinf$) with objects restricted to natural numbers $n$ seen as finite sets of cardinality $n$.

This view of distribution monads via Lawvere theories has been explored by various authors (see e.g.~\cite{furber-jacobs-calco13,fritz-presentation,bonchi-sobocinski-zanasi-cmcs16,jacobs-distribution-monads}). We prove that $\LawvConv$ and $\LawvSConv$ have all finite coproducts, adopting the view of Kleisli maps as stochastic matrices, where the Kleisli composition corresponds in this context to matrix multiplication. This approach is also present in~\cite{fritz-presentation}.

\begin{lem}%
\label{lem:B1:LawvConv-finite-products}
The categories $\LawvConv$ and $\LawvSConv$
have all finite products.
\end{lem}

\begin{proof}
We show that the Lawvere theories $\LawvConv$ and $\LawvSConv$ have all finite products (with addition as product) by showing that the Kleisli categories $\kl_\N(\DMinf)$ and $\kl_\N(\SDMinf)$ have all finite coproducts (with addition as coproduct).

For every natural number $n \in \N$, there is exactly one stochastic matrix of size $n \times 0$
and therefore $0$ is an initial object for $\kl_\N(\DMinfopt)$.

Identity maps are defined to be $\eta_n:n \to \DMinfopt(n)$. We call the corresponding $n \times n$ stochastic matrix $1_n$ and consider the inclusion maps
$\kappa_1:n_1 \to n_1 + n_2$
and
$\kappa_2:n_2 \to n_1 + n_2$
as the stochastic matrices
$K_1=\left(\begin{smallmatrix}
           1_{n_1}\\
           0_{n_2}
\end{smallmatrix}\right)$
and
$K_2=\left(\begin{smallmatrix}
           0_{n_1}\\
           1_{n_2}
\end{smallmatrix}\right)$.

Now, consider a pair of stochastic matrices $A_1$ and $A_2$, with corresponding maps $f_1:n_1 \to p$ and $f_2:n_2 \to p$ (with $n_1,n_2,p \in \N$).

To satisfy the universal property of the coproduct, we must construct a unique map $f:n_1 + n_2\to p$ such that the equation $f_i = f \circ \kappa_2$ holds for $i \in \{1,2\}$. Then, we observe that the stochastic matrix
$A=\left(\begin{smallmatrix}
           A_1 & A_2
\end{smallmatrix}\right)$ is the unique stochastic matrix whose multiplication by $K_i$ gives $A_i$ (for $i \in \{1,2\}$) and therefore, we define $f$ to be the Kleisli map corresponding to the stochastic matrix $A$.
%
\end{proof}

Then, the coproduct $f_1+f_2:n_1+n_2\to p_1 + p_2$ of two Kleisli maps $f_1:n_1 \to p_1$ and $f_2:n_2 \to p_2$ is defined as the diagonal
$A_1 + A_2
\defeq
\left(
\begin{smallmatrix}
A_1 & 0\\
0 & A_2
\end{smallmatrix}
\right)
$ of their corresponding stochastic maps $A_1$ and $A_2$.
It follows that $\LawvConv$ and $\LawvSConv$ are Lawvere theories, since they are strict monoidal categories when one consider $+:\LawvOptConv \times \LawvOptConv \to \LawvOptConv$ as tensor product, with the natural number $0$ as unit.

\medskip
Recall that the category $\DcpoS$ of pointed dcpos and strict Scott-continuous maps is monoidal closed when equipped with the smash product defined in the introduction. Now, observe that the Lawvere theory $\LawvSConv$ is a small $\DcpoS$-category: for every pair $(n,m)$ of natural numbers, the homset
\[
\LawvSConv(n,m)
\defeq
{\SDMinf(n)}^m
\]
is a dcpo as a finite product of dcpos. Indeed, the set $\SDMinf(X)$ is known to be a dcpo when equipped with the pointwise order~\cite{hasuo-jacobs-sokolova}:
\[
\varphi \leq \psi \iff \forall x. \varphi(x) \leq \psi(x)
\]
In fact, one can observe that the coproduct functor $+:\LawvOptConv \times \LawvOptConv \to \LawvOptConv$ is a $\DcpoS$-enriched functor, turning the category $\LawvSConv$ into a small symmetric monoidal $\DcpoS$-enriched category $(\LawvSConv,+,0)$.

In what follows, we make use of the following notions.

\begin{defi}
An endofunctor $F$ on a $\Dcpo_{\perp !}$-enriched category $\cat{C}$ is locally continuous, locally monotone, and locally strict if \[F_{X,Y} : \cat{C}(X,Y) \to \cat{C}(FX,FY)\] is Scott-continuous, monotone, and strict, respectively.
\end{defi}

It turns out that Kegelspitzen are models of this Lawvere theory $\LawvSConv$, as explained in the following theorem. In essence, this theorem represents Kegelspitzen as domain-theoretic stochastic matrices.

\begin{thm}%
\label{th:KS-LawvSConv-DcpoS}
The category $\KS$ of Kegelspitzen and affine Scott-continuous maps is equivalent to
the category ${[\LawvSConv,\DcpoS]}_\times$ of models of the $\DcpoS$-enriched Lawvere theory $\LawvSConv$ of subconvex sets,
i.e.~the category of finite product-preserving locally strict Scott-continuous functors $\LawvSConv \to \DcpoS$ and natural transformations between them.
\end{thm}

\begin{proof}
Recall that Kegelspitzen can be equivalently defined as dcpos $X$ with Scott-continuous maps $X^n \to X$  and a product ${(x_i)}_{1 \leq i \leq n} \in X^n$ as the convex sum $\sum_i r_i \cdot x_i \in X$ for $r \in \LawvSConv(n,1)$, one can define a functor $\Phi: \KS \to {[\LawvSConv,\DcpoS]}_\times$ which acts as follows on objects:
\begin{align*}
\Phi(X)(n) &= X^n\quad (n \in \N)\\
\Phi(X)(r: n \to 1)({(x_i)}_i) &= \sum_i r_i \cdot x_i
\end{align*}

So any Kegelspitze $X$ can be identified with a (finite) product-preserving functor $\Phi(X):\LawvSConv \to \DcpoS$, i.e.~a model of the Lawvere theory $\LawvConv$ in the category $\DcpoS$, defined as follows. For $n \in \N$, $\Phi(X)(n)=X^n \in \DcpoS$.

A function $r: n \to 1$ is a $n$-ary operation definable in the Lawvere theory $\LawvSConv$ of subconvex sets,
and as such it induces a function $f_r:X^n \to X$,  defined by
\[
f_r(x_1,\ldots,x_n)=\sum_i r_i \cdot x_i
\]
which is Scott-continuous in each argument
since $X$ is taken to be a Kegelspitze.
Consequently, the function $f_r:X^n\to X$ is taken to be $\Phi(X)(r):\Phi(X)(n)\to\Phi(X)(1)$.

Then the mapping $\Phi$ can be turned into a functor $\Phi: \KS \to {[\LawvSConv, \DcpoS]}_\times$
which acts as follows on maps:
an affine Scott-continuous map $f: X \to Y$ is associated to a natural family of strict Scott-continuous maps $\Phi(f): \Phi(X) \nattrans \Phi(Y)$,
where ${\Phi(f)}_n:X^n \to Y^n$ is the strict Scott-continuous map
\[
f^n: {(x_i)}_{1 \leq i \leq n} \mapsto {(f(x_i))}_{1 \leq i \leq n}
\]
for every $n \in \N$.

The faithfulness of the functor $\Phi$ is entailed by its construction:
\[
\forall f, g \in \KS(X,Y). (\Phi(f)=\Phi(g) \implies f={\Phi(f)}_1={\Phi(g)}_1=g)
\]
Additionally, we are required to prove that the functor $\Phi$ is full. Consider a natural transformation $\alpha:\Phi(X) \nattrans \Phi(Y)$ for some Kegelspitzen $X$ and $Y$. In what follows we show that there is an affine strict Scott-continuous map $f$ such that $\alpha = \Phi(f)$.

By construction, the strict Scott-continuous map $f \defeq \alpha_1: X \to Y$ induces the whole natural transformation $\alpha$,
i.e.~$\alpha_n = f^n$ for every $n \in \N$. Indeed,
from the commuting square
\[
\xymatrix@R+.5pc{
n \ar[d]_-{\delta_i} 
& X^n \ar[r]^-{\alpha_n} \ar[d]_-{\Phi(X)(\delta_i)} 
& Y^n \ar[d]^-{\Phi(Y)(\delta_i)}\\ 
1
& X \ar[r]_-{f} 
& Y
}
\]
where $1 \leq i \leq n$ and $\delta_i$ is the Dirac notation introduced in Definition~\ref{def:B1:distribution-monads},
we deduce that for every $1 \leq i \leq n$ and for $x=(x_1,\ldots,x_n) \in X^n$,
\[
f(x_i)=f(\Phi(X)(\delta_i)(x))=\Phi(Y)(\delta_i)(\alpha_n(x))={(\alpha_n(x))}_i
\]
Moreover, the strict Scott-continuous map $\alpha_1:X \to Y$ is affine, i.e.~is a morphism in $\KS$:
this is entailed by the commuting square
\[
\xymatrix@R+.5pc{
n \ar[d]_-{r} 
& X^n \ar[r]^-{\alpha_n} \ar[d]_-{\Phi(X)(r)} 
& Y^n \ar[d]^-{\Phi(Y)(r)}\\ 
1
& X \ar[r]^-{\alpha_1} 
& Y
}
\]
where $r \in \LawvSConv(n,1)$, which means that
\[
\forall x=(x_1,\ldots,x_n)\in X^n. \alpha_1(\sum_i r_i \cdot x_i) = \sum_i r_i \cdot {(\alpha_n(x))}_i
\]
i.e.
\[
\forall x=(x_1,\ldots,x_n)\in X^n. \alpha_1(\sum_i r_i \cdot x_i) = \sum_i r_i \cdot \alpha_1(x_i)
\]

This concludes our proof that the functor $\Phi$ is full, since $\alpha_n=f^n=\Phi(f)(n)$ for every $n \in \N$, and therefore $\alpha=\Phi(f)$.  The full and faithful functor $\Phi$ turns out to be essentially surjective, and therefore an equivalence: a model $F: \LawvSConv \to \DcpoS$ is equivalent to the model $\Phi(X)$, where $X$ is the Kegelspitze formed by the dcpo $F(1)$ together with the Scott-continuous convex structure $F(\LawvConv(n,1))$.
%
%
\end{proof}

It is worth noting that using a similar reasoning, one can show that the category $\Conv$ of convex sets and affine maps is equivalent to the category ${[\LawvConv,\Set]}_\times$ of models of the Lawvere theory $\LawvConv$ of convex sets, and that the category $\dConv$ of convex dcpos and Scott-continuous affine maps is equivalent to the category ${[\LawvConv,\Dcpo]}_\times$ of models of the Lawvere theory $\LawvConv$ of convex sets in the category $\Dcpo$ of dcpos and Scott-continuous maps. Those observations along with Theorem~\ref{th:KS-LawvSConv-DcpoS} can be seen as instances of the standard result (see e.g.~\cite{jacobs-coalgebra-book}) that the Eilenberg Moore category $\mathcal{EM}(T)$ of a monad $T$ is equivalent to the category ${[\opp{\kl_\N(T)},\Set]}_\times$, since we have the following chain of equivalences
\[
\SConv \cong \mathcal{EM}(\SDMinf) \cong {[\opp{\kl_\N(\SDMinf)},\Set]}_\times \cong {[\Lawv,\Set]}_\times
\]

\newcommand{\LawvCone}{\mathbb{L}_\text{Cone}}
\newcommand{\RR}{\mathbb{R}}
A similar categorical construction can be obtained for cones, here defined as $\RR^+$ semi-rings, which also have their order-theoretic counterparts.

\begin{defi}
An \textit{ordered cone} $C$ is a cone equipped with a partial order $\leq$ such that addition and scalar multiplication are monotone. That is, $a \leq b$ implies that $a+c \leq b+c$ and $r \cdot a \leq r \cdot b$, for every $a,b,c \in C$ and every $r \in \R^+$.

An ordered cone $A$ is a \textit{d-cone} (resp.~a \textit{b-cone}) when its order is directed-complete (resp.~bounded directed-complete), and its addition $+: A \times A \to A$ and its scalar multiplication $\cdot: \unit \times A \to A$ are Scott-continuous maps.  We write $\cat{dCone}$ (resp. $\cat{bCone}$) for the category of d-cones and (resp.\ b-cones) with Scott-continuous maps.
\end{defi}

We refer the interested reader to~\cite{keimel-plotkin-kegelspitzen} for a thorough study of those domain-theoretic structures.

\newcommand{\MMon}{\mathcal{M}}
In this setting, the Lawvere theory of cones $\LawvCone$ is defined with the multiset monad $\MMon$ on the semiring $\R^+$ which acts as follows on objects
\[
\MMon(X)=\left\{~\varphi:X \to \R^+ ~\middle|~ \text{supp}(\varphi)
\text{ finite}~\right\}
\qquad
\text{ where }
\quad
\text{supp}(\varphi)=\left\{~x \in X ~\middle|~ \varphi(x)\neq 0~ \right\}
\]
In other words, the Lawvere theory of cones $\LawvCone$ is the category of natural numbers together with functions $n \to m$ seen as Kleisli maps $m \to \MMon(n)$, i.e.~$\LawvCone$ is the opposite category $\opp{\kl_\N(\MMon)}$ of the restricted Kleisli category of the multiset monad $\MMon$. Replaying every step of our reasoning with the multiset monad instead of the distribution monad leaves us with the following equivalences:
\[
\cat{dCone} \cong {[\LawvCone,\Dcpo]}_\times
 \qquad \qquad
\cat{bCone} \cong {[\LawvCone,\cat{BDcpo}]}_\times
\]
In other words, d-cones are models of the Lawvere theory of cones in the category of dcpos and Scott-continuous maps, while b-cones are models of the Lawvere theory of cones in the category of bdcpos~\cite{abramsky-jung} and Scott-continuous maps.

\medskip
Last but not least: the isomorphism between the categories $\cat{KS}$ and ${[\LawvSConv,\DcpoS]}_\times$ establishes a formal relation between the category $\KS$ and the category $\DcpoS$, which is known to be symmetric monoidal closed when equipped with the smash product $\otimes_\perp$, with its internal hom $\KS(-,-)$ as exponential (see e.g.~\cite[Section~1.3]{jung-cartesian}).

\newcommand{\sarrow}{\overset{\perp}{\arrow}}
\begin{prop}%
\label{prop:B1:KS-mon-closed}
The category $\cat{KS}$ is monoidal closed with respect to the smash product $\otimes_\perp$
and the internal hom functor $\KS(-,-)$
\end{prop}

\newcommand{\smashprod}{\otimes_\perp}
\begin{proof}
As the smash product of two pointed (convex) dcpos, the smash product of two Kegelspitzen is a pointed convex dcpo whose convex structure is defined as follows: a convex sum in the set $X \smashprod Y$ is given by
$\sum_{i \leq n,j \leq m} \lambda_i \gamma_j (x_i, y_j) \defeq (\sum_{i \leq n} \lambda_i x_i, \sum_{j \leq m} \gamma_j y_j)$
with vectors $\vec{\lambda}=(\lambda_1,\ldots,\lambda_n)$ and $\vec{\gamma}=(\gamma_1,\ldots,\gamma_m)$ of non-negative real numbers (respectively associated to ${(x_i)}_i$ and ${(y_j)}_j$) such that $\sum_{1 \leq i \leq n} \lambda_i = 1$ and $\sum_{1 \leq j \leq n} \gamma_j = 1$, and therefore $\sum_{i,j} \lambda_i \gamma_j = 1$.

Now, we observe that for every pair $(X,Y)$ of Kegelspitzen, the set $\KS(X,Y)$ is convex when equipped with a convex structure defined pointwise on the convex structure of the Kegelspitze $Y$. The least upper bound $\lub_i f_i$ of a directed set ${\{f_i \}}_{i \in I}$ of strict Scott-continuous functions between Kegelspitzen is also strict Scott-continuous. It remains to show that when every $f_i$ ($i \in I$) is affine, so does $\lub_i f_i$ since $Y$ is a Kegelspitzen and therefore $\theta_{n,Y}:\SDMinf(n) \times Y \to Y$ is affine in both coordinates:
\begin{align*} 
 (\lub_i f_i)(\sum_{1 \leq j \leq n} r_j \cdot x_j)
 &= \lub_i (f_i(\sum_j r_j \cdot x_j))\\
 &= \lub_i (\sum r_j \cdot f_i(x_j))\\
 &= \lub_i \theta_{n,Y} ((r_j)_{j \leq n},(f_i(x_j))_{j \leq n})\\ 
 &= \theta_{n,X} ((r_j)_{j \leq n},(\lub_i(f_i(x_j)))_{j \leq n})\\ 
 &= \sum_j r_j \cdot (\lub_i f_i)(x_j)
\end{align*}
for every convex sum $\sum_{1 \leq j \leq n} r_j \cdot x_j$ in the Kegelspitze $X$.

Therefore, $\KS(X,Y)$ is a pointed convex dcpo, which satisfies the Kegelspitzen condition since $Y$ does:
\begin{align*}
\forall \lambda \in \unit. \forall x \in X.\quad
(\lambda \cdot (\lub_i f_i))(x)
&= \lambda \cdot ((\lub_i f_i)(x)) = \lambda \cdot (\lub_i f_i(x)) \\
&= \lub_i \lambda \cdot f_i(x) = \lub_i (\lambda \cdot f_i)(x)\\
&= (\lub_i (\lambda \cdot f_i))(x)
\end{align*}

\newcommand{\ev}{\text{ev}}
Moreover, the strict Scott-continuous evaluation map $\ev_{X,Y}:\KS(X,Y) \otimes_\perp X \to Y$, given by the monoidal closed structure of $\DcpoS$~\cite[Section~1.3]{jung-cartesian}, is affine:
\begin{align*}
\ev_{X,Y}\left(\sum_{i \leq n,j \leq m} \lambda_i \gamma_j \cdot (f_i, x_j)\right)
&= \ev_{X,Y}\left(\sum_i \lambda_i \cdot f_i, \sum_j \gamma_j \cdot x_j\right)\\
&= \sum_i \lambda_i \cdot f_i(\sum_j \gamma_j \cdot x_j)\\
&= \sum_{i,j} \lambda_i \gamma_j \cdot (f_i(x_j))\\
&= \sum_{i,j} \lambda_i \gamma_j  \cdot (\ev_{X,Y}(f_i,x_j))
\end{align*}
for every convex sum $\sum_{i \leq n,j \leq m} \lambda_i \gamma_j (f_i, x_j)$ in the Kegelspitzen $\KS(X,Y) \otimes_\perp X$.

Finally, the curryfied form $\Lambda(f):X \to \KS(Y,Z):x\mapsto f(x,-)$ of an affine strict Scott-continuous map $f: X \otimes_\perp Y \to Z$ is also strict Scott-continuous~\cite[Section~1.3]{jung-cartesian} and affine, since one can verify that for every convex sum $\sum_i r_i \cdot x_i \in X$ and every $y \in Y$,
\[
\Lambda(f)(\sum_i r_i \cdot x_i)(y) = \sum_i r_i \cdot \Lambda(f)(x_i)(y)
\]
This concludes our proof that we have, for every triplet $(X,Y,Z)$ of Kegelspitzen, the following bijective correspondence in $\KS$:
\newcommand{\stimes}{\otimes_\perp}
\[
\begin{prooftree}
\qquad f: X \stimes Y \to Z \qquad
\Justifies
\qquad \Lambda(f): X \to \KS(Y,Z) \qquad
\end{prooftree}
\]
for which the equation $\ev_{X,Y} \circ (\Lambda(f) \stimes \id_X) = f$ holds.
\end{proof}

From the observation that every full subcategory of the cartesian closed category $\Dcpo$ which contains the singleton dcpo, the cartesian product $\times$ and the exponential $\arrow \defeq \Dcpo(-,-)$ is itself cartesian closed~\cite{jung-cartesian}, we obtain the following proposition.

\newcommand{\KSScott}{\KS_{\text{Scott}}}
\begin{prop}%
\label{prop:B1:KS-cart-closed}
The category {$\KSScott$}
of Kegelspitzen and Scott-continuous maps is cartesian closed.
\end{prop}

Note that in the category $\KSScott$, maps between Kegelspitzen are not necessarily affine, and in particular do not necessarily preserve least elements.

\section{Interpreting pPCF}%
\label{sec:B1:interpreting-pPCF}

In this section, we consider a probabilistic extension of PCF~\cite{plotkin-PCF}, named pPCF\footnote{The presentation of this language essentially follows the work of Ehrhard et al., see e.g.~\cite{ehrhard-pagani-tasson-arXiv}}, whose types and terms are defined as follows:
\begin{align*}
 \text{Types: } t, u, \ldots &::= \text{nat} \mid t \arrow u\\
 \text{Terms: } M, N, \ldots &::= \pcfnum{n} \mid x \mid \tsucc(M) \mid \text{if}(M, P, z \cdot Q) \mid \lambda x^t. M \mid (M) N \mid \tcoin(\kappa) \mid \tfix(M)
\end{align*}
where $n \in \N$, $x, y, \ldots$ are symbols for variables and $\kappa \in \unit \cap \QQ$ is a probability. We associate those grammars to the following typing rules.

\[
\begin{prooftree}
\justifies
\Gamma, x:t \vdash x:t
\end{prooftree}
\qquad
\begin{prooftree}
\Gamma, x:t \vdash M:u
\justifies
\Gamma \vdash \lambda x^t. M:t \arrow u
\end{prooftree}
\qquad
\begin{prooftree}
\Gamma \vdash M: t \arrow u \quad \Gamma \vdash N:t
\justifies
\Gamma \vdash (M)N:u
\end{prooftree}
\qquad
\begin{prooftree}
\Gamma \vdash M: t\arrow t
\justifies
\Gamma \vdash \tfix(M):t
\end{prooftree}
\]
\[
\begin{prooftree}
\justifies
\Gamma \vdash \pcfnum{n}:\tnat
\end{prooftree}
\qquad
\begin{prooftree}
\Gamma \vdash M:\tnat
\justifies
\Gamma \vdash \tsucc(M):\tnat
\end{prooftree}
\qquad
\begin{prooftree}
\kappa \in \unit \cap \QQ
\justifies
\Gamma \vdash \tcoin(\kappa):\tnat
\end{prooftree}
\]
\[
\begin{prooftree}
\Gamma \vdash M:\tnat \quad \Gamma \vdash P:t \quad \Gamma, z:\tnat \vdash Q:t
\justifies
\Gamma \vdash \tif(M, P, z \cdot Q):t
\end{prooftree}
\]

The associated reduction transition is probabilistic: terms $\tcoin(\kappa)$ reduce to $\pcfnum{0}$ with probability $\kappa$ and to $\pcfnum{1}$ with probability $1-\kappa$. This construction is associated to the following reduction rules.

\[
\begin{prooftree}
\justifies
\tcoin(\kappa) \probred \pcfnum{0}
\end{prooftree}
\qquad
\begin{prooftree}
\justifies
\tcoin(\kappa) \xprobred{1-\kappa} \pcfnum{1}
\end{prooftree}
\qquad
\begin{prooftree}
M \probred N
\justifies
\tsucc(M) \probred \tsucc(N)
\end{prooftree}
\]

We write $\detred$ for deterministic reductions, i.e.~probabilistic reductions $\probred$ with $\kappa=1$. The deterministic reduction $\detred$ allows us to reuse standard reduction rules, that is:
\[
\begin{prooftree}
M \detred N
\justifies
M \xprobred{1} N
\end{prooftree}
\qquad
\begin{prooftree}
\justifies
\tfix(M) \detred (M)\tfix(M)
\end{prooftree}
\qquad
\begin{prooftree}
\justifies
\tsucc(\pcfnum{n}) \detred \pcfnum{n+1}
\end{prooftree}
\]
Context rules are given as follows:
\[
\begin{prooftree}
\justifies
(\lambda x^t. M)N \detred M[x \mapsto N]
\end{prooftree}
\qquad
\begin{prooftree}
M \probred N
\justifies
(M)P \probred (N)P
\end{prooftree}
\]
Note that such context rules means that the probabilistic reduction  $\probred$ is a weak-head reduction: the leftmost outermost redex is reduced first, and there is no reduction under abstraction.~\footnote{This approach is consistent with the presentation of pPCF in~\cite{ehrhard-pagani-tasson-arXiv}.}

We amend the traditional if-then-else instruction $\text{if}(M, P, Q)$ in order to prevent the loss of the value $\pcfnum{n}$ obtained from the evaluation of the term $M$: when $M$ reduces to $\pcfnum{0}$, one can evaluate $P$ knowing that $n=0$ but when $M$ reduces to $\pcfnum{n+1}$ $(n \in \N)$, it is necessary to associate a variable $z = \pcfnum{n}$ in order for the term $Q$ to reuse the value of $n$. This leads to conditional constructions $\tif(M,P, z \cdot Q)$ associated to the following reduction rules which adopt a call-by-value strategy on the ground type $\tnat$, in the sense that the term $M:\tnat$ is evaluated first, and the resulting value is used for conditional branching.

\[
\begin{prooftree}
\justifies
\tif(\pcfnum{0}, P, z \cdot Q) \detred P
\end{prooftree}
\qquad
\qquad
\begin{prooftree}
\justifies
\tif(\pcfnum{n+1}, P, z \cdot Q) \detred Q[z \mapsto \pcfnum{n}]
\end{prooftree}
\]
\[
\begin{prooftree}
M \probred N
\justifies
\tif(M,P, z \cdot Q) \probred \tif(N,P, z \cdot Q)
\end{prooftree}
\]

\medskip
By construction, for every judgement $\Gamma \vdash M: t$, the judgement $\Gamma \vdash M': t$ holds whenever $M \probred M'$ holds.

\begin{lem}[Substitution Lemma]
Suppose that $\Gamma, x: u \vdash M: t$ and $\Gamma \vdash P: u$.
If $M \detred M'$ then $M[x \mapsto P] \detred M'[x \mapsto P]$.
\end{lem}

\begin{proof}
This lemma can be proven by induction on terms.
Terms which apply a term to another are the non-trivial cases of this proof.

Consider a term $M = (N) L$, when $N$ isn't an abstraction and reduces to another term $N'$. Then, the reduction $N \detred N'$ implies that there is a reduction
\[
M = (N) L \detred (N') L
\]
and since $M \detred M'$ by hypothesis, we have that $M' = (N')L$.

First, let us observe that $N$ cannot be a variable since $N \detred N'$. Now, assuming that $\Gamma \vdash P: u$, one can deduce that $N[x \mapsto P]$ is not an abstraction since $N$ isn't, and finally by induction hypothesis,  $N[x\mapsto P] \detred N'[x\mapsto P]$ and therefore:
\[
((N)L)[x\mapsto P]=(N[x \mapsto P])L[x\mapsto P] \detred (N'[x \mapsto P])L[x \mapsto P] = ((N')L)[x\mapsto P]
\qedhere
\]
\end{proof}

This extension of PCF allows to define the predecessor of a term $M$ by:
\[
\text{pred}(M) \defeq \lambda x^\tnat.~\tif(x, 0, z \cdot z)
\]
Moreover, probabilistic combinations of terms $M:t$ and $N:t$ under the probability $\kappa$ are given by the term:
\[
M \oplus_\kappa N \defeq \tif(\tcoin(\kappa),M,N)
\]
The language allows a manipulation of (first-order) probabilistic data (of type $\tnat$) through a let construction which corresponds to a probabilistic programming perspective to sampling:
\newcommand{\tlet}{\text{let}}
\newcommand{\tin}{\text{in}}
\[
\tlet\, x = M\, \tin\, N \defeq \tif(M,N[x \mapsto \pcfnum{0}], z \cdot N[x \mapsto \tsucc(z)])
\]

It is possible to give an interpretation to this language in the cartesian closed category $\KSScott$ of Kegelspitzen and Scott-continuous maps. In short, types $t$ can be interpreted as Kegelspitzen $\denot{t}$, contexts $\Gamma = (x_1:t_1, \ldots, x_n:t_n)$ as Kegelspitzen $\denot{t_1} \times \cdots \times \denot{t_n}$, and terms $\Gamma \vdash M:t$ as Scott-continuous maps $\denot{\Gamma \vdash M:t}:\denot{\Gamma} \to \denot{t}$, with the following denotations:
\[
\denot{\nat} = \SDMinf(\N)
\qquad \text{ and } \qquad
\denot{t \arrow u} = \denot{t} \arrow \denot{u} \defeq \Dcpo(\denot{t},\denot{u})
\]
In what follows, functions $\varphi:\N\to\unit$ in $\SDMinf(\N)$ are written as sequences ${(\varphi(n))}_{n \in \N}$. In particular, since closed terms $\vdash M:\tnat$ are interpreted by functions $\denot{\vdash M:\tnat}:\N\to\unit$ in $\SDMinf(\N)$, we write $\denot{M:\tnat}_n$ for $\denot{\vdash M:\tnat}(n)$.

\[
\denot{\Gamma \vdash x_i:t_i}=\pi_i:\rho\mapsto \rho_i
\qquad
\denot{\Gamma \vdash \num{0}:\tnat}(\rho)=(1,0,\cdots)
\]
\[
\denot{\Gamma \vdash \tcoin(\kappa):\tnat}(\rho) = \kappa \cdot \denot{\Gamma \vdash \num{0}}(\rho) + (1-\kappa) \cdot \denot{\Gamma \vdash \num{1}}(\rho)
\]
\[
\denot{\Gamma \vdash \tsucc(M):\tnat}(\rho)=(0,u_0, u_1,\cdots)
\qquad
\text{where } u=\denot{\Gamma \vdash M:\tnat}(\rho)
\]
\[
\denot{\Gamma \vdash \tif(M,P, z \cdot Q):t}(\rho)=v_0 u + \sum_{i \geq 1} v_i u'_i
\]
where $v=\denot{\Gamma \vdash M:\nat}(\rho)$, $u=\denot{\Gamma \vdash P:t}(\rho)$, and for $i \geq 1$,
\[
u'_i=\denot{\Gamma, z:\tnat \vdash Q:t}(\rho,e_{i-1})
= \denot{\Gamma \vdash Q[z \mapsto \underline{i-1}]:t}
\]
(with $e_i$ defined to be the ``non-probabilistic'' integer $i$ with weight $1$ for $i$ and weight $0$ for each $j \neq i$).
\[
\denot{\Gamma \vdash \tfix(M):t}(\rho)=\fixop(\denot{\Gamma \vdash M:t \arrow t}(\rho))
\text{ where }
\fixop(f)=\lub_{n}f^n(\perp)
\]
\[
\denot{\Gamma \vdash (M)N:t}(\rho)=f(x)
\text{ where } f=\denot{\Gamma \vdash M:u \arrow t}(\rho) \text{, } x=\denot{\Gamma \vdash N:u}(\rho)
\]
\[
\denot{\Gamma \vdash \lambda x^u.M: u \arrow t}(\rho)(x)=\denot{\Gamma,x:u \vdash M:t}(\rho,x)
\]

One of the interesting properties of this denotational semantics is that the interpretation of a term can be expressed as a sum of the interpretations of the terms it reduces to.

\begin{lem}[Invariance of the interpretation]%
\label{lem:B1:invariance}
Suppose that the judgement $\Gamma \vdash M: t$ holds, for some term $M$ which isn't a value (i.e.~normal form). Then, the following equality holds
\[
\denot{\Gamma \vdash M: t}
= \sum_{M \probred M'} \kappa \cdot \denot{\Gamma \vdash M': t}
\]
\end{lem}

\begin{proof}
We first consider the case of judgements $\Gamma \vdash M:t$ such that the term $M$ reduce through the deterministic reduction rules: if $M \detred M'$, then the interpretations of the terms that we have just defined ensures that $\denot{\Gamma \vdash M} = \denot{\Gamma \vdash M'}$. For example, for the judgement $\Gamma \vdash (\lambda x^t.M)N:u$ (with $x:t$ and $N:t$) such that $(\lambda x^t.M)N \detred M[x \mapsto N]$, we have
\[
\denot{\Gamma \vdash (\lambda x^t.M)N:u}(\rho)=\denot{\Gamma, x:t \vdash N:u}(\rho,\denot{\Gamma\vdash N:t}(\rho))=\denot{\Gamma \vdash M[x \mapsto N]}(\rho)
\]

It remains to show that the terms which reduce through probabilistic reduction rules (with $\kappa < 1$) satisfy the invariance property. By the construction of our reduction system, such terms are of the form $\tcoin(\kappa)$, $(M)P$, $\tif(M,P,z\cdot Q)$, or $\tsucc(M)$. We now show that the invariance property is satisfied in those four cases.

First, let us observe that the interpretation of $\tcoin(\kappa):\tnat$ under any context $\Gamma$ can be re-written as follows:
\[
\denot{\Gamma \vdash \tcoin(\kappa):\tnat}(\rho) = \sum_{M \probred \pcfnum{n}} \kappa \cdot \denot{\Gamma \vdash \pcfnum{n}:\nat}
\]

For the remaining three cases, we proceed by induction on judgements.
Consider terms $\tsucc(M):\tnat$ and $\tif(M,P,z\cdot Q)$ (where $M \neq \pcfnum{n}$ for some $n \in \N$, with $P:t$ and $Q:t$), and $(N)P:t$ (with $P:u$) such that the judgements $\Gamma \vdash M:\tnat$ and $\Gamma \vdash N:u \arrow t$ satisfy the invariance property. From our operational semantics, we deduce that if $\tsucc(M)\probred Q$, then $Q$ is of the form $\tsucc(M')$ for some term $M':\tnat$ such that $M \probred M'$.
Similarly, if $\tif(M,P,z\cdot Q) \probred R$ then $R$ is of the form $\tif(M',P,z\cdot Q)$ for some term $M':\tnat$ such that $M \probred M'$, and
if $(N)P \probred Q$ then $Q$ is of the form $(N')P$ for some term $N':u \arrow t$ such that $N \probred N'$. And since by induction hypothesis, we have
\[
\denot{\Gamma \vdash M} = \sum_{M \probred M'} \denot{\Gamma \vdash M':\tnat}
\qquad
\text{ and }
\qquad
\denot{\Gamma \vdash N} = \sum_{N \probred N'} \denot{\Gamma \vdash N':u \arrow t}
\]
then we have by the construction of our denotational semantics the following equalities:
\[
\denot{\Gamma \vdash \tsucc(M)}
= \sum_{\tsucc(M) \probred \tsucc(M')} \denot{\Gamma \vdash \tsucc(M'):\tnat}
= \sum_{\tsucc(M) \probred Q} \denot{\Gamma \vdash Q:\tnat}
\]
\[
\denot{\Gamma \vdash \tif(M,P, z \cdot Q):t}(\rho)
= \sum_{\tif(M,P, z \cdot Q) \probred \tif(M',P, z \cdot Q)} \denot{\Gamma \vdash \tif(M',P, z \cdot Q):t}(\rho)
\]
\[
\denot{\Gamma \vdash (N)P:t}
= \sum_{(N)P \probred (N')P} \denot{\Gamma \vdash (N')P:t}
= \sum_{(N)P \probred Q} \denot{\Gamma \vdash Q:t}
\qedhere
\]
\end{proof}

In line with similar approaches~\cite{danos-ehrhard-PPCF,ehrhard-tasson-pagani-PCS-FA-pPCF}, the probabilities of the transitions of pPCF terms can be organised as follows (see~\cite[Sec.~1.2]{ehrhard-pagani-tasson-arXiv}).

\begin{defiC}[{\cite[Section~1.2]{ehrhard-pagani-tasson-arXiv}}]%
\label{def:B1:PPCF-terms-Markov}
In what follows, we write $\Lambda$ for the set of all pPCF terms and we say that a term $M$ is \emph{weak-normal} when there is no probabilistic reduction $M \probred M'$.
The \emph{matrix of pPCF terms} is the stochastic matrix $\Prob \in \unit^{\Lambda \times \Lambda}$ defined by
\[\Prob_{M,M'}=\begin{cases}
   \kappa \text{ if } M \probred M' \\
   1 \text{ if } M=M' \text{ is weak-normal}\\
   0 \text{ otherwise}
  \end{cases}
  \]
\end{defiC}

Using Definition~\ref{def:B1:PPCF-terms-Markov}, we formulate the following soundness property which is a restatement of Lemma~\ref{lem:B1:invariance}, which established the invariance of interpretation. In this context,

\begin{prop}[Soundness]%
\label{prop:B1:soundness}
Suppose that the judgement $\Gamma \vdash M: t$ holds, for some term $M$ which isn't a value. Then, the following equality holds
\[
\denot{\Gamma \vdash M: t} = \sum_{M' \text{ term}} \Prob_{M,M'} \cdot \denot{\Gamma \vdash M':t}
\]
\end{prop}

By applying repeatedly this lemma and considering the specific case of normal forms, one obtains the following corollary.

\begin{cor}%
\label{cor:B1:step-towards-adequacy}
Consider a closed type $\vdash t$.
For $\Gamma \vdash M: t$ and $k \in \N$, the following equality holds
\[\denot{\Gamma \vdash M: t} = \sum_{M' \text{ term}} \Prob^k_{M,M'} \denot{\Gamma \vdash M':t}.\]
where $\Prob^k_{M,M'}$ is the probability that the term $M$ reduces to the term $M'$ in $k$ steps.

Then for every closed term $\vdash M: \tnat$, we have the inequality
\[
\denot{M:\tnat}_n \geq \Prob^\infty_{M,\pcfnum{n}}
\text{ where }
\Prob^\infty_{M,\pcfnum{n}}
\defeq \sup_k~(\Prob^k_{M,\pcfnum{n}})
\]
i.e.~where $\Prob^\infty_{M,\pcfnum{n}}$ is the least upper bound of the probabilities that $M$ reduced to $\pcfnum{n}$ in finitely  many steps.
\end{cor}

\begin{proof}
Applying Proposition~\ref{prop:B1:soundness}, we have:
\[
\denot{M:\tnat}_n = \sum_{M':\tnat} \Prob_{M,M'} \cdot \denot{M':\tnat}_n
\geq \Prob^\infty_{M,\pcfnum{n}} \cdot \denot{\pcfnum{n}}_n
= \Prob^\infty_{M,\pcfnum{n}} \cdot 1
= \Prob^\infty_{M,\pcfnum{n}}
\qedhere
\]
\end{proof}

\section{Computational adequacy}%
\label{sec:B1:adequacy}

In this section, we provide a computational adequacy result (for the type $\tnat$), that is we prove the converse of the inequality expressed in Corollary~\ref{cor:B1:step-towards-adequacy}, which is:
\[
\forall \vdash M:\tnat, \quad \denot{M: \tnat}_n \leq \Prob^\infty_{M,\pcfnum{n}}
\]

The key to the proof of this inequality is to define a logical relation, taken from~\cite{danos-ehrhard-PPCF} but inspired by the original article on the semantics of PCF~\cite{plotkin-PCF}.

\begin{defi}
For every type $t$, consider the relation $\triangleleft_t \subseteq \denot{t} \times \Lambda_t$ between the denotation $\denot{t}$ and the set $\Lambda_t$ of all closed terms of type $t$, written with an infix notation and defined by induction as follows:
\[
x = {(x_n)}_{n \in \N} \triangleleft_{\nat} M \equiv \forall n. x_n \leq \Prob^\infty_{M,\num{n}}
\]
\[
f \triangleleft_{u \arrow t} M \equiv \forall x. \forall \vdash P: u. (x \triangleleft_u P \implies f(x) \triangleleft_t (M)P )
\]
Note that once again, we follow the convention of presenting elements of $\SDMinf(\N)$ as sequences ${(x_n)}_{n \in \N}$.
\end{defi}

This logical relation has the following closure properties.

\begin{lem}[Closure properties of the logical relation]
Consider $\vdash M: t$
\begin{enumerate}
 \item\label{item:closure1} If $\vdash M: t$ and $M \detred M'$, then $x \triangleleft_t M$ holds if and only if $x \triangleleft_t M'$ holds;
 \item\label{item:closure2} $0 \triangleleft_t M$ holds;
 \item\label{item:closure3} $\sup_n x_n \triangleleft_t M$ holds for every increasing sequence ${(x_n)}_n$ in $\denot{t}$ such that $x_n \triangleleft_t M$ for $n \in \N$;
 \item\label{item:closure4} $x_0 \cdot y + (\sum_i x_{i+1}) \cdot z \triangleleft_\tnat \ifop(M,P,z \cdot Q)$ holds for $x, y, z \in \denot{\tnat}$ and $\vdash M: \tnat, \vdash P: \tnat, \vdash Q: \tnat$ such that $x \triangleleft_\tnat M, y \triangleleft_\tnat P, z \triangleleft_\tnat Q$.
\end{enumerate}
\end{lem}

\begin{proof}
The closure property (\ref{item:closure2}) follows from the fact that probabilities are positive numbers, while the closure property (\ref{item:closure3}) follows from the fact that Scott-continuous functions are ordered pointwise.

As for the closure property (\ref{item:closure4}), we first observe that if the term $\tif(M,P,z \cdot Q)$ reduces to $\num{n}$ for some $n\in \N$, then either $M$ reduces to $\num{0}$ and $P$ reduces to $\num{n}$, or $M$ reduces to $\num{n+1}$ (for some $n\in \N$) and $Q$ reduces to $\num{n}$. Then, the closure property (\ref{item:closure4}) is induced by the following equation which is valid for every $n\in \N$ (see~\cite[Lemma~38]{danos-ehrhard-PPCF}):
\[
\Prob^\infty_{\tif(M,P,z \cdot Q),\num{n}} =
\Prob^\infty_{M,\num{0}} \cdot \Prob^\infty_{P,\num{n}}
+ \sum_{k \geq 0} \Prob^\infty_{M,\num{k+1}} \cdot \Prob^\infty_{Q,\num{n}}
\]

Now, we proceed by induction to obtain a proof of the closure property (\ref{item:closure1}). When $t=\tnat$, the property is straightforward from the observation that $\Prob^k_{M',\num{n}}=\Prob^{k+1}_{M,\num{n}}$. Let us now consider the case in which $t = u \arrow v$.

Assume that $f \triangleleft_t M$. When $M$ isn't an abstraction, $(M)P \detred (M')P$ for every closed term $P$ of type $u$, and we can apply the definition of the logical relation:
\[
\forall \vdash P:u, x \in \denot{u}, x \triangleleft_u P \xRightarrow{f \triangleleft_t M} f(x) \triangleleft_v (M)P \xRightarrow{\text{induction hypothesis}} f(x) \triangleleft_v (M')P
\]
When $M$ is an abstraction $\lambda x^u. N:v$ with $x:u \vdash N:v$, there is a term $N'$ such that $N \detred N'$. Then by the Substitution Lemma,
\[
(M)P \detred N[x \mapsto P] \detred N'[x \mapsto P]
\]
and therefore we obtain $f(x) \triangleleft_v N'[x \mapsto P]$ by applying the induction hypothesis twice. Hence, since $(M')P \detred N'[x \mapsto P]$, we have $f(x) \triangleleft_v M'(P)$ by induction, which concludes our proof that $f \triangleleft_t M'$.

Conversely, assume $f \triangleleft_t M'$. We focus on the case in which $M$ is an abstraction $\lambda x^u.N:v$ with $x:u \vdash N:v$ (since the case in which $M$ isn't an abstraction is again trivial). Then for every closed term $\vdash P:u$ and every $x \in \denot{u}$, we have $f \triangleleft_t \lambda x^u.N$ and therefore
\[
f(x) \triangleleft_v (\lambda x.N')P \detred N'[x \mapsto P]
\]
therefore $f(x) \triangleleft_v N'[x \mapsto P]$ (again by the substitution lemma and the induction hypothesis). Then, we have $f(x) \triangleleft_v N[x \mapsto P]$ and by induction $f(x) \triangleleft_v (M)P=(\lambda x^u. N)P$ since $(\lambda x^u. N)P \detred N[x \mapsto P]$.
\end{proof}

Using the closure properties of the logical relation, we prove the following lemma by induction.

\begin{lem}
Consider a judgment $\Gamma \vdash M:u$ where $\Gamma \equiv (x_1:t_1, \ldots, x_n:t_n)$.

$\denot{\Gamma \vdash M:u}(\rho) \triangleleft_u M[P/x]$, every family  $P={\{P_i\}}_{1 \leq i \leq n}$ of closed terms of type ${\{t_i\}}_{1 \leq i \leq n}$ (i.e.~$\vdash P_i: t_i$) and every family $x={\{x_i\}}_{1 \leq i \leq n}$ of variables of type $t={\{t_i\}}_{1 \leq i \leq n}$ such that $\denot{\Gamma \vdash x_i:t_i}(\rho) \triangleleft_{t_i} P_i$.
\end{lem}

\begin{proof}
We will reason by induction on terms.

Case $M=x_i$: $\denot{\Gamma \vdash x_i:t_i}(\rho) \triangleleft_{t_i} P_i=x_i[P/x]$

Case $M=\num{l}$: there is only one transition path $\num{l} \to \num{l}$ of probability $1$ and length $0$.

Case $M=\tsucc(N)$: straightforward induction.

Case $M=\ifop(N,L,R)$: follows from the closure property of the logical relation for $\ifop$.

Case $M=\tcoin(\kappa)$: There is exactly one transition path to $\pcfnum{0}$ with probability $\kappa$, and one transition path to $\pcfnum{1}$ with probability $1-\kappa$. It follows that
\[
\Prob^\infty_{\tcoin(\kappa),\pcfnum{0}} = \kappa
\text{ and }
\Prob^\infty_{\tcoin(\kappa),\pcfnum{1}} = 1-\kappa
\]
We write:
\[
\denot{\Gamma \vdash \tcoin(\kappa):\tnat}(\rho)
= \Prob^\infty_{\tcoin(\kappa),\pcfnum{0}}
\cdot \denot{\Gamma \vdash \pcfnum{0}:\tnat}(\rho)
+
\Prob^\infty_{\tcoin(\kappa),\pcfnum{1}}
\cdot \denot{\Gamma \vdash \pcfnum{1}:\tnat}(\rho)
\]
and therefore
\[
\denot{\Gamma \vdash \tcoin(\kappa):\tnat}(\rho)(n) = \Prob^\infty_{\tcoin(\kappa),\pcfnum{n}}
\]
for every $n \in \N$,
i.e.
\[
\denot{\Gamma \vdash \tcoin(\kappa):\tnat}(\rho) \triangleleft \tcoin(\kappa)=\tcoin(\kappa)[x \mapsto P]
\]

Case $M=(N)L$: straightforward induction, based on the definition of the logical relation $\triangleleft_{t \arrow u}$ on the type $t \arrow u$.

Case $M=\lambda y^t.N:t \arrow u$: Given any element $y \in \denot{t}$ and any closed term $Q$ of type $t$ such that $y \triangleleft_t Q$, we have that
\[
\denot{\Gamma \vdash \lambda y. N}(\rho)=\denot{\Gamma, x:t \vdash N}(\rho,x) \triangleleft_{u} N[P/x,Q/y]
\]
by induction hypothesis. Then
\[
\denot{\Gamma \vdash \lambda x. N}(\rho)(y) \triangleleft_u (\lambda y^t.N[P/x])Q
\]
by the closure property of the logical relation for the deterministic reduction
\[
(\lambda y^t.N[P/x])Q \detred N[P/x,Q/y]
\]

Case $M=\fixop(N)$ with $\Gamma \vdash N: u \arrow u$: the function
\[
f \defeq \denot{\Gamma \vdash N}(\rho):\denot{u} \to \denot{u}
\]
is a Scott-continuous function such that
\[
\denot{\Gamma \vdash M}(\rho)=\lub_k f^k(\perp)
\]
Then, by the closure property of the logical relation for fixpoints, it suffices to prove by induction on $k$ that
\[
f^k(\perp) \triangleleft_u \fixop(N[P/x])
\]
for every $k \in \N$, knowing that the property already holds for $k=0$.

Suppose that $f^k(\perp) \triangleleft_u \fixop(N')$, where $N'=N[P/x]$, for some $k \in \N$. By our induction hypothesis (on terms),
\[
f \triangleleft_{u \arrow u} N'=N[P/x]
\qquad
\text{ and thus }
\qquad
f^{k+1}(\perp) \triangleleft_u N'\fixop(N')
\]

Finally, one can conclude that $f^{k+1}(\perp) \triangleleft_u N'$ by observing that $\fixop(N') \detred N'\fixop(N')$ and applying the closure property of the logical relation for deterministic transitions.
\end{proof}

This lemma provides us an adequacy theorem.

\begin{thm}[Computational adequacy]%
\label{thm:B1:adequacy}
For every closed term $M$ of type $\tnat$,
\[
\denot{M: \tnat}_n = \Prob^\infty_{M,\num{n}}
\]
\end{thm}

\begin{proof}
For every closed term $\vdash M: \tnat$, we have proven previously that
\[
\denot{M: \tnat}_n \geq \Prob^\infty_{M,\num{n}}
\qquad
\text{ and thus }
\qquad
\denot{M: \tnat}_n = \Prob^\infty_{M,\num{n}}
\]
since by the adequacy lemma, $\denot{M: \tnat} \triangleleft_\nat M$, i.e.
$\denot{M}_n \leq \Prob^\infty_{M,\num{n}}$ for every $n \in \N$.
\end{proof}

We just provided a computationally adequate model for pPCF, alternative to probabilistic coherence spaces (see e.g.~\cite{ehrhard-tasson-pagani-PCS-FA-pPCF}).
Although the type $\tnat$ has the same denotation in the two semantics,
the resemblance between the two semantical models is lost at higher types. Although our adequacy theorem is formulated in a similar fashion as in~\cite{ehrhard-tasson-pagani-PCS-FA-pPCF},
it is unclear to us whether there exists an interesting categorical relation between Kegelspitzen and probabilistic coherence spaces.
\hide{A probabilistic coherence space is given by a pair $(A,P(A))$ of a countable set and a carrier $P(A)$ with several coherence properties which ensure that it is a dcpo with respect to the pointwise order on real functionals $A \to \R^+$ but does not ensure that it constitute a subset of $\SDMinf(A) \defeq \left\{ v \in {(\R^+)}^A \mid \sum_a v_a \leq 1\right\}$.}

\section{Interpreting recursive types}%
\label{sec:B1:recursive-types}

In this section, we discuss the interpretation of recursive types, taking as a basis their formalization in the language FPC\@. Note that, unlike PCF, FPC is a linear language (without contraction). Moreover, in spite of its lack of fixed-point operator (unlike PCF), general recursion can be implemented in FPC and the language is Turing-complete (as is PCF). We detail a presentation of FPC in Appendix~\ref{app:FPC}. But first, let us pause for a moment and recall some categorical notions which are essential in the interpretation of languages such as FPC, which cater for recursive types.

\subsection{Involutory category theory}%
\label{sub:B1:involutory}

As a preliminary to the description of the denotation of recursive types with Kegelspitzen, we recall briefly here Fiore's ``Doubling Trick''~\cite[Section~6.3]{fiore-thesis} (also mentioned in~\cite[Section~4.2.3]{mccusker-semantics-recursive-types}), an universal categorical construction which allows to turn mixed-variance functors $\opp{\cat{C}} \times \cat{C} \to \cat{D}$ into covariant functors $\opp{\cat{C}} \times \cat{C} \to \opp{\cat{D}} \times \cat{D}$. This property is required because the denotation of recursive types requires to be able to find fixpoints, not only for covariant (endo)functors but also for mixed-variance functors. Indeed, the arrow functor $\cdot \arrow \cdot:\opp{\KS} \times \KS \to \KS$ is a mixed variance functor. 

In what follows, the category $\symcat{\cat C}$
is short for $\opp{\cat{C}} \times \cat{C}$. Additionally, in categories with binary products $\times$, we write
\[
f_1 \defeq \pi_1 \circ f: X \to Y_1
\qquad
\text{ and }
\qquad
f_2 \defeq \pi_2 \circ f: X \to Y_2
\]
for the composite of the morphism $f \in \cat{C}(X, Y_1 \times Y_2)$.

\begin{defiC}[{\cite[Definition~4.6]{fiore-FPC}}]
An \emph{involutory category} is the pair $(\cat{C},\text{Inv}_C)$ of a locally small category $\cat{C}$ together with an \emph{involution functor} $\text{Inv}_C:\cat{C}\to\opp{\cat{C}}$, i.e.~a functor $\text{Inv}_C:\cat{C}\to\opp{\cat{C}}$ such that
$\opp{(\text{Inv}_C)} \circ \text{Inv}_C = \mathrm{Id}_C$, the identify functor on the category $\cat{C}$.

We write {$\cat{InvCat}$} for the large cartesian category of involutory categories and homomorphisms
\[
F:(\cat{C},\text{Inv}_C)\to(\cat{D},\text{Inv}_D)
\]
defined as functors $F:\cat{C}\to\cat{D}$ such that
\[
\opp{F} \circ \text{Inv}_C = \text{Inv}_D \circ F
\]
\end{defiC}

A canonical example is the pair $(\symcat{\cat C},\text{Swap}_C)$ where $\text{Swap}_C \defeq \langle \Pi_2, \Pi_1 \rangle$ (with $\Pi_1$, $\Pi_2$ projections given by the cartesian structure).

\begin{defi}
A functor $F:\symcat{\cat C} \to \symcat{\cat D}$ is \emph{symmetric} if $F:(\symcat{\cat C},\text{Swap}_C)\to(\symcat{\cat D},\text{Swap}_D)$ is a morphism in $\cat{InvCat}$, i.e.
\[
F_1(f,g) = F_2(g,f)
\text{ for maps }
f
\text{ in the category }
\opp{\cat{C}}
\text{ and }
g
\text{ in the category }
\cat{C}
\]
\end{defi}

It turns out that mixed-variance functors induce symmetric functors, and every symmetric functor arises in that way, following a result due to Fiore in~\cite[Section~4.4]{fiore-FPC}, re-proven by McCusker in~\cite[Section~4.2.3]{mccusker-semantics-recursive-types}.

\begin{prop}%
\label{prop:B1:corr-mixed-co}
There is a one-to-one correspondence
\[
\begin{prooftree}
F: \symcat{\cat C} \to \cat{D}
\Justifies
\symcat{F}: \symcat{\cat C} \to \symcat{\cat D}
\end{prooftree}
\]
between mixed variance functors $F: \symcat{\cat C} \to \cat{D}$ and symmetric functors $\symcat{F}: \symcat{\cat C} \to \symcat{\cat D}$ defined by
\begin{align*}
\symcat{F}(A,B) \defeq (F(B,A),F(A,B))
\qquad \qquad
\symcat{F}(f,g) \defeq (F(g,f),F(f,g))
\end{align*}
In particular, for every B\'enabou cosmos $\cat V$, the functor $\symcat{F}$ is $\cat V$-enriched whenever the categories
$\symcat{\cat C}$ and $\cat{D}$,
and the functor $F$ are $\cat V$-enriched.
\end{prop}

\subsection{Algebraic compactness of the category of Kegelspitzen}%
\label{sub:B1:algebraic-compactness}

One of the issues with the inclusion of recursive types in a probabilistic language such as pPCF
is that
the cardinality of $\denot{t \arrow u}$ might be strictly larger than that of $\denot{t}$ in some cases, which might prevent the existence of a fixpoint for $t$.
Exploiting the presentation of the category $\KS$ as a category of models of the Lawvere theory of subconvex sets, we re-use the notion of algebraic compactness, which guarantees the existence of such fixpoints.

\medskip
Recall that a category $\cat{C}$ is algebraically compact for a class $\mathcal{L}$ of endofunctors on $\cat{C}$ if every endofunctor $F$ in the class $\mathcal{L}$ has a canonical fixpoint $\mu F$, which is the initial $F$-algebra and at the same time the inverse of the final $F$-coalgebra.

\medskip
To obtain the algebraic compactness of $\KS$ for locally continuous endofunctors, we rely on the notion of colimits of $\omega$-chains, and the following theorem.

\begin{defi}
An $\omega$-\emph{chain} in a category $\cat{C}$ is a sequence of the form\ $\Delta = D_0 \xrightarrow{\alpha_0} D_1 \xrightarrow{\alpha_1} \cdots$

Given an object $D$ in a category $\cat{C}$, a \emph{cocone} $\mu: \Delta \to D$ for the $\omega$-chain $\Delta$ is a sequence of arrows (commonly referred to as \textit{embeddings}) $\mu_n : D_n \to D$ such that  $\mu_n = \mu_{n+1} \circ \alpha_n$ holds for every $n \geq 0$.

A colimit (or colimiting cocone) of the $\omega$-chain $\Delta$ is an initial cocone from $\Delta$ to $D$, i.e.~for every cocone $\mu' : \Delta \to D'$, there exists a unique map $f : D \to D'$ such that $f \circ \mu_n = \mu'_n$ holds for every $n \geq 0$.

Dually, we consider $\omega^{\operatorname{op}}$-chains $\Delta^{\operatorname{op}} = D_0 \xleftarrow{\beta_0} D_1 \leftarrow \cdots$  in a category, cones $\gamma: \Delta^{\operatorname{op}} \leftarrow D$ and limits (or limiting cones) for an $\omega^{\operatorname{op}}$-chain $\Delta^{\operatorname{op}}$.
\end{defi}

\begin{thm}%
\label{thm:alg-compact-DcpoS-enriched}
For every small $\DcpoS$-category $\cat{C}$, the $\DcpoS$-enriched category of locally strict continuous functors $\cat{C} \to \DcpoS$ and natural transformations between them (ordered pointwise) is algebraically compact for the class of locally continuous endofunctors.
\end{thm}

\begin{proof}
First we need to show that $[\cat C, \DcpoS]$ has all colimits of $\omega$-chains of embeddings.

Consider an endomorphism $\Psi$ on the category $[\cat C,\DcpoS]$. Consider an $\omega$-chain $\Delta = {(F_k,\alpha_k:F_k \nattrans F_{k+1})}_{k \in \N}$ in $[\cat C, \DcpoS]$, where $F_0$ is the zero functor of $[\cat C,\DcpoS]$ and $\alpha_0$ is the unique natural transformation
from $F_0$ to $\Psi F_0$, and $F_{k+1}=\Psi F_k$, $\alpha_{k+1}=\Psi \alpha_k$ for $k \in \N$. Then $\Delta$ is a family of $\omega$-chains $\Delta_X = {(F_k(X), \alpha_k[X]:F_k(X)\to F_{k+1}(X))}_{k \in \N}$ in $\DcpoS$ indexed by the objects of the small category $\cat C$.

For each $X \in \cat C$, the corresponding $\omega$-chain $\Delta_X$ in $\DcpoS$ has a colimit, since $\DcpoS$ has all colimits of $\omega$-chains of embeddings. For each $X \in \cat C$, we write $\mu[X]:\Delta_X \to F(X)$ for the cocone (of arrows $\mu_k[X]:F_k(X) \to F(X)$) which is the colimit of the $\omega$-chain $\Delta_X$.

We turn the mapping $F:X\mapsto F(X)$ into a functor $F:\cat C \to \DcpoS$ by taking
\[
F(f)=\mu_0[X'] \circ F_0(f) \circ \mu_0^P[X]
\]
for $f \in \cat C(X,X')$. The definition of $F(f)$ for each $f \in \cat C(X,X')$ is such that for every $k \in \N$, the following diagram commutes:

\[
\xymatrix@R+.5pc{
F_k(X) \ar[d]_-{F_k(f)} \ar[r]^{\mu_k[X]} 
& F(X) \ar[r]^-{\mu^P_0[X]} \ar[d]_-{F(f)} 
& F_0(X) \ar[d]^-{F_0(f)}\\ 
F_k(X') \ar[r]^{\mu_k[X']} 
& F(X') \ar[r]^-{\mu^P_0[X']} 
& F_0(X')
}
\]
so that the equality $F(f) = \mu_k[X'] \circ F_k(f) \circ \mu_k^P[X]$ holds for every $k \in \N$.

Consider the family of cocones $\Delta_X \to F(X)$ (whose arrows are the embeddings $\mu_k[X]:F_k(X) \to F(X)$, $k \in \N$) indexed by the objects of the category $\cat C$. It is a cocone $\Delta \to F$ whose arrows are embeddings  $\mu_k: F_k \nattrans F$, which are well-defined natural transformation by construction of $F$.
It follows that $\Delta \to F$ is the colimit of $\Delta$ because for each $X \in \cat C$, $\Delta_X \to F(X)$ is the colimit of $\Delta_X$.

Consider the category $\omega$-$\cat{CPO}$ of $\omega$-cpos, i.e.\ posets in which every countable chain has a sup (which is the case for dcpos), and Scott-continuous maps between them. A category which is $\DcpoS$-enriched is also $\omega$-$\cat{CPO}$-enriched and a functor which preserves directed sups preserves in particular countable directed sups. We know from~\cite[Theorem~5.4]{barr-algebraically} that every $\omega$-$\cat{CPO}$-enriched category which has colimits of $\omega$-chains is algebraically compact for the class of locally continuous endofunctors, and therefore so does $[\cat C, \DcpoS]$.
\end{proof}

Recall that the Lawvere theory $\LawvSConv$ is a small $\DcpoS$-category. Then, the fact that the functor category $[\LawvSConv,\DcpoS]$ is algebraically compact for locally continuous endofunctors leads us to the following theorem.

\begin{cor}%
\label{cor:B1:KS-alg-compt}
The category $\KS$, as a category equivalent to the category ${[\LawvSConv,\DcpoS]}_\times$, is algebraically compact for locally continuous endofunctors.
\end{cor}

\begin{proof}
First, let us observe that every locally continuous endofunctor $F$ on ${[\LawvSConv,\DcpoS]}_\times$ extends to a locally continuous endofunctor $G$ on $[\LawvSConv,\DcpoS]$ defined by $G(X)=F(X)$ when $X:\LawvSConv \to \DcpoS$ is product-preserving, and $G(X)=X$ otherwise.

Now, consider a chain of embeddings  ${(D_n,\alpha_n:D_n \nattrans D_{n+1})}_n$ formed of product-preser\-ving functors $\LawvSConv \to \DcpoS$ and natural families of strict Scott-continuous maps, where
\[
D_0 \defeq 1:\LawvSConv\to\DcpoS
\qquad
\text{ and }
\qquad
D_{n+1} \defeq G(D_n) = F(D_n)
\text{ for }n \in \N
\]
By Theorem~\ref{thm:alg-compact-DcpoS-enriched}, we know that the functor $G$ has a fixpoint $D:\LawvSConv\to\DcpoS$ given on objects by
\[
D(k) = \{ {(x_n)}_n \in \Pi_n D_n(k) \mid \forall n \geq 0, \alpha_n^P(k)(x_{n+1})=x_n\}
\]
where every $\alpha_n^P:D_{n+1}\nattrans D_n$ is part of an embedding projection pair $\eppair{\alpha_n^E}{\alpha_n^P}$, with $\alpha_n^E \defeq \alpha_n$, as described in~\cite[\S~5.2]{fiore-FPC}.
Since every functor $D_n:\LawvSConv\to\DcpoS$ is product-preserving, so is $D$: for natural numbers $k$ and $l$, we have
\begin{align*}
 D(k+l) &= \{ {(x_n)}_n \in \Pi_n D_n(k+l) \mid \forall n \geq 0, \alpha_n^P(k+l)(x_{n+1})=x_n\}\\
 &\cong \{ ({(y_n)}_n,{(z_n)}_n) \in \Pi_n D_n(k) \otimes \Pi_n D_n(l) \mid \forall n \geq 0, (\alpha_n^P(k)(y_{n+1})=y_n \\
 &\wedge  \alpha_n^P(l)(z_{n+1})=z_n)\}\\
 &\cong D(k) \otimes D(l)
\end{align*}
It follows that $F(D)$ is equal to $G(D)$, which is itself equivalent to $D$.
\end{proof}

The denotational semantics of types introduced in Section~\ref{sub:B1:KS-FPC} essentially relies on the category $\abs{\KS} \defeq \opp{\KS} \times \KS$. The algebraic compactness of $\abs{\KS}$ can be obtained through standard results of the literature~\cite{adamek-algebraically,barr-algebraically,freyd-algebraically}, gathered in~\cite{fiore-FPC}:
\begin{itemize}
 \item Algebraic compactness is a self-dual property: if the category $\cat{C}$ is algebraically compact for locally continuous endofunctors, then so does its opposite category $\opp{\cat C}$.
 \item If the categories $\cat{C}$ and $\cat{D}$ are algebraically compact for locally continuous endofunctors, then so does their product category $\cat{C} \times \cat{D}$.
\end{itemize}

\begin{cor}
The category $\abs{\KS}$ is algebraically compact for locally continuous endofunctors.
\end{cor}

\subsection{Kegelspitzen as a model of recursive types}%
\label{sub:B1:KS-FPC}

As an algebraically compact category, the category $\KS$ is a domain-theoretic model of recursive types. We recall here the foundations of the semantics of recursive types, and refer the interested reader to Fiore's thesis~\cite{fiore-thesis} for a complete account of the axiomatization of computationally adequate models of FPC\footnote{Note that the FPC defined in Appendix~\ref{app:FPC} is a linear calculus. In the typing rule for the application $(M)N$, the terms $M$ and $N$ are given distinct contexts and therefore do not share variables. We motivate this restriction by the fact that the category $\KS$, in which we have chosen to interpret recursive types, is not a cartesian closed category.}

Type judgements $\Theta \vdash t$ and judgements $\Theta \mid \Gamma \vdash M:t$ (introduced in Appendix~\ref{app:FPC}) are respectively denoted by symmetric locally Scott-continuous $n$-ary functors
\[
\denot{\Theta \vdash t}: \symcat{\KS}^n \to \symcat{\KS}
\]
and by natural transformations
\[
\denot{\Theta \mid \Gamma \vdash M:t}:\denot{\Theta \vdash \Gamma} \nattrans \denot{\Theta \vdash t}
\]
i.e.~natural families of morphisms
\[
\big\{
\denot{\Theta \mid \Gamma \vdash M:t}_X:
\denot{\Theta \vdash \Gamma}(X) \to \denot{\Theta \vdash t}(X)
\mid X \in \symcat{\KS}^n
\big\}
\]
in the category $\symcat{\KS}$.

The denotation $\denot{\Theta \vdash \Theta_i}$ of the type judgement $\Theta \vdash \Theta_i$ (with $\Theta$ typing context of length $n$) is the $i$-th projection functor $\Pi^{\symcat{\KS}^n}_i: \symcat{\KS}^n \to \symcat{\KS}$. Moreover, the denotation $\denot{\Theta \vdash \mu X. t}$ of a typing judgement $\Theta \vdash \mu X.t$ involving a recursive type $\mu X.t$ to be $\mu \denot{\Theta, X \vdash t}$, the fixpoint of the functor $\denot{\Theta, X \vdash t}:\symcat{\KS}^{n+1} \to \symcat{\KS}$ by algebraic compactness.

Now, recall that for functors $F,G: \symcat{C} \to \symcat{D}$,
we have functors $\Pi^{\symcat{\cat{C}}}_2 F, \Pi^{\symcat{\cat{C}}}_2 G:\symcat{\cat{C}} \to \cat{D}$,
and therefore a (mixed-variance) functor $\Pi^{\symcat{\cat{C}}}_2 F \otimes \Pi^{\symcat{\cat{C}}}_2 G:\symcat{\cat{C}}\to \cat{D}$,
itself in one-to-one correspondence with a symmetric functor $\symcat{\Pi^{\symcat{\cat{C}}}_2 F \otimes \Pi^{\symcat{\cat{C}}}_2 G}:\symcat{C}\to\symcat{D}$ by Proposition~\ref{prop:B1:corr-mixed-co}. Then, the denotations of other type contexts is given as follows.
\[
\denot{\Theta \vdash t_1 \times t_2}
\defeq
\symcat{\Pi^{\symcat{\cat{\KS}}}_2\denot{\Theta \vdash t_1} \otimes_\perp \Pi^{\symcat{\cat{\KS}}}_2\denot{\Theta \vdash t_2}}
\]
\[
\denot{\Theta \vdash t_1 + t_2}
\defeq
\symcat{\Pi^{\symcat{\cat{\KS}}}_2\denot{\Theta \vdash t_1} \oplus \Pi^{\symcat{\cat{\KS}}}_2\denot{\Theta \vdash t_2}}
\]
\[
\denot{\Theta \vdash t_1 \arrow t_2}
\defeq
\symcat{\KS(\Pi^{\symcat{\cat{\KS}}}_1\denot{\Theta \vdash t_1},
\Pi^{\symcat{\cat{\KS}}}_2\denot{\Theta \vdash t_2})}
\]

where $\Pi^{\symcat{\cat{\KS}}}_1: \symcat{\cat{\KS}} \to \opp{\cat{\KS}}$
and $\Pi^{\symcat{\cat{\KS}}}_2: \symcat{\cat{\KS}} \to \cat{\KS}$ are the projections
of the cartesian product $\symcat{\cat{\KS}}$, $\otimes_\perp:\cat{\KS} \times \cat{KS} \to \cat{KS}$ is the smash product functor, $\KS(-,-):\opp{\cat{\KS}} \times \cat{KS} \to \cat{KS}$ is the homset functor (which acts as exponential object in the monoidal closed structure $(\KS,\otimes_\perp,\KS(-,-))$ of Proposition~\ref{prop:B1:KS-mon-closed}). Note that function types are interpreted as Scott-continuous functions in pPCF, while they are here interpreted as affine (strict) Scott-continuous maps.

The functor $\oplus:\cat{\KS} \times \cat{KS} \to \cat{KS}$ is the functor induced by the coproduct of convex sets, discussed in a categorical setting in~\cite{jacobs-westerbaan-fossacs15} and adapted for (pointed) convex dcpos in~\cite[Section~3.1.2]{rennela-staton-mfps15}.

In detail, recall that the sum $A+B$ of two convex sets, $A$ and $B$,
can be described as the set
$A\uplus B\uplus (A\times B\times (0,1))$,
where $(0,1)$ is the open unit interval.
Its elements either come directly from $A$, or from $B$, or
are a non-trivial formal convex combination of elements from $A$ and $B$.
With a slightly informal notation, we write $(a,-,0)$ instead of $a$, and $(-,b,1)$ instead of $b$.
Then define the convex structure as follows
\[\sum_{i}r_i.(a_i,b_i,\lambda_i)\stackrel{\text{def}}=
(\sum_i\frac{r_i(1-\lambda_i)}{1-\sum_i r_i\lambda_i}.a_i,\sum_i\frac{r_i\lambda_i}{\sum_i r_i\lambda_i}.b_i,({\textstyle \sum_i r_i\lambda_i}))\]
taking the obvious convention where $({\textstyle \sum_i r_i\lambda_i})$ is $0$ or $1$.
This has the universal property of the coproduct in the category of convex sets.
\newcommand{\ssum}{\oplus}
Therefore, if $A$ and $B$ are (sub)convex dcpos then 
we define their \emph{skew sum} $A\ssum B$ as
the coproduct $A+B$ of $A$ and $B$ as convex sets, equipped with the partial order
$(a,b,\lambda)\leq (a',b',\mu)$ if
$a\leq a'$ and $b\leq b' $ and $\lambda\leq \mu$. In which case, $A \ssum B$ is a Kegelspitze when $A$ and $B$ are Kegelspitzen.

It is worth noting that this has a universal property similar to the universal property of a coproduct, to the exception that there is
an additional requirement that $a\leq b$ for $a\in A$, $b\in B$.
For example, we can freely add a bottom element to a convex dcpo $A$ by
taking the skew sum $(1\ssum A)$. Moreover, the skew sum is not a symmetric operator, unlike its syntactic counterpart (the sum type) in FPC, as the skew sum is constrained to account for the fact that pFPC manipulates probability distributions.

\subsubsection*{Acknowledgements}
I would like to thank Bart Jacobs, Robin Kaarsgaard, Ohad Kammar, Klaus Keimel, Michael Mislove, Michele Pagani, Christine Tasson and Fabio Zanasi for helpful discussions, and more particularly Sam Staton for suggesting the problem. Some of the research leading to the results of the present work was undertaken while the author was based at Radboud University, and funded by the
European Research Council under the European Union's Seventh Framework
Programme (FP7/2007-2013) / ERC grant agreement n. 320571. The author's visits to the Department of Computer Science of Oxford University have been financially supported by a Royal Society University Research Fellowship. 

\bibliographystyle{plain}
\bibliography{lmcs-pfpc-v2}

\appendix
\section{FPC}%
\label{app:FPC}
The functional programming language FPC~\cite{fiore-FPC}
can be seen as a ``PCF with recursive types'', and has been heavily used in the denotational study of recursive types. A recursive type is an inductively defined data type for terms which may contain type variables that are used in fixed points. It is an important concept for high-level programming languages, which allows the definition of data types such as the types for lists and trees, whose size can dynamically grow. An example of recursive type in a ML-style functional programming language is
\begin{lstlisting}
type nat = zero | succ nat
\end{lstlisting}
which corresponds to the natural numbers.

In recursive type theory, recursive types are written $\mu X.t$, where $X$ is a type variable which may appear in the type $t$. For example, the type nat is written $\mu X. 1 + X$. Indeed, the constructor zero is a type without arguments and therefore corresponds to the unit type $1$, and succ takes as argument another term of type nat.

The syntax of FPC relies on two grammars, one for types and one for terms:
\begin{align*}
 \text{Types }t, u &::= X \mid t + u \mid t \times u \mid t \to u \mid \mu X.t\\
\text{Terms }M, N, P &::= x \mid \op{inl}_{t,u}(M) \mid \op{inr}_{t,u}(M) \mid \opcase{M}{N}{P} \mid (M)N\\
&\mid (M,N) \mid \lambda x^\sigma.M \mid \op{fst}(M) \mid \op{snd}(M) \mid \op{intro}_{\mu X. t}(M) \mid \op{elim}(M)
\end{align*}

where $X$ is taken in the sort of type variables, and $x$ is taken in the sort of variables. In detail, we have sum types $t + u$, product types $t \times u$, function types $t \to u$, and recursive types {$\mu X.t$}, and corresponding primitives to manipulate instances of such types. In particular, instructions such as $\op{intro}_{\mu X. t}(M)$ and $\op{elim}(M)$ allow respectively the introduction and the elimination of recursive types, through a process that we now proceed to describe.

Firstly, we need to define the rules which describe well-formed types and expressions. For that purpose, we introduce typing judgements $\Theta \vdash t$, which indicate that the type $t$ is a \textit{well-formed type} with respect to the \textit{typing context} $\Theta$. This means that the free variables of the type $t$ are in the list $\Theta$ of distinct type variables. Recall that a variable is called \textit{free} when it is not bound. In this setting, a type variable is free when it is not used as a parameter of a recursive type. For example, the variable $X$ is bound in $\mu X.t$ for every type $t$. A closed type is a well-formed type with no typing context, that is a type $t$ such that the typing judgement $\vdash t$ holds. The substitution in a type $t$ of every occurence of a type variable $X$ by a type $t'$ is written {$t[X \mapsto t']$}. Well-formed types of FPC are defined inductively by the following rules:

\[
\begin{prooftree}
\justifies
\Theta, X \vdash X
\end{prooftree}
\qquad \qquad
\begin{prooftree}
\Theta, X \vdash t
\justifies
\Theta \vdash \mu X. t
\end{prooftree}
\qquad \qquad
\begin{prooftree}
\Theta \vdash t \quad \Theta \vdash u
\justifies
\Theta \vdash t \star u
\using
\star \in \{+,\times,\to\}
\end{prooftree}
\]

Similarly, one can define well-formed expressions inductively, using judgements {$\Theta \mid \Gamma \vdash M:t$} which entails that the type $t$ is well-formed under the typing context $\Theta$ (that is, the typing judgement $\Theta \vdash t$ holds), and that the term $M$ of type $t$ is well-formed under the context $\Gamma$, defined as a list of distinct variables, written as $x:t$. What follows is a set of rules which allows to determine inductively which expressions are well-formed:

\[
\begin{prooftree}
\Theta \mid \Gamma \vdash M:t[X \mapsto \mu X. t]
\justifies
\Theta \mid \Gamma \vdash \textbf{intro}_{\mu X. t}(M):\mu X.t
\end{prooftree}
\qquad
\begin{prooftree}
\Theta \mid \Gamma \vdash M:\mu X. t
\justifies
\Theta \mid \Gamma \vdash \textbf{elim}(M): t[X \mapsto \mu X. t]
\end{prooftree}
\]
\medskip
\[
\begin{prooftree}
\justifies
\Theta \mid \Gamma, x: t \vdash x:t
\end{prooftree}
\quad \quad
\begin{prooftree}
\Theta \mid \Gamma, x: t \vdash M:u
\justifies
\Theta \mid \Gamma \vdash \lambda x^t. M:t \to u
\end{prooftree}
\quad \quad
\begin{prooftree}
\Theta \mid \Gamma \vdash M:t \to u \quad \Theta \mid \Gamma' \vdash N: t
\justifies
\Theta \mid \Gamma,\Gamma' \vdash (M)N:u
\end{prooftree}
\]
\medskip
\[
\begin{prooftree}
\Theta \mid \Gamma \vdash M:t \quad \Theta \vdash u
\justifies
\Theta \mid \Gamma \vdash \textbf{inl}_{t,u}(M):t+u
\end{prooftree}
\quad \quad
\begin{prooftree}
\Theta \mid \Gamma \vdash M:t \quad \Theta \vdash u
\justifies
\Theta \mid \Gamma \vdash \textbf{inr}_{t,u}(M):u+t
\end{prooftree}
\]
\medskip
\[
\begin{prooftree}
\Theta \mid \Gamma \vdash M:t+u \quad \Theta \mid \Gamma', x:t \vdash N:v \quad \Theta \mid \Gamma', y:u \vdash P:v
\justifies
\Theta \mid \Gamma,\Gamma' \vdash \opcase{M}{N}{P}:v
\end{prooftree}
\]

\medskip
Now, we can define a \textit{program} in FPC to be an expression $M$ such that the judgement $\vdash M:t$ holds for some type $\vdash t$, that is: $M$ is a closed term of closed type. A \textit{context with a hole} is an expression
{$C[-]$} with holes such that for every term $M$, $C[M]$ is the expression obtained by replacing every hole by the term $M$. When the context $C[-]$ is of type $t$, we write $C[-]:t$.

\medskip
Secondly, the grammars of FPC are associated with the following \textit{operational semantics}, which describes how programs are executed. But first, let's recall what a reduction system is.

\begin{defi}
A \textit{reduction system} is a pair $(\Lambda,\to)$ of a collection $\Lambda$ of terms and a binary relation $\to \subseteq \Lambda \times \Lambda$ on terms, which is called a \textit{reduction relation}. The transitive reflexive closure of a reduction relation $\to$ is denoted by $\to^*$. And therefore, if the relation
{$M \to N$} means that the term $M$ reduces to the term $N$ in one step, then the relation
{$M \to^* N'$} means that the term $M$ reduces to the term $N$ in finitely many steps. A term $M \in \Lambda$ is a \textit{normal form} (or \textit{value}) if there is no term $N \in \Lambda$ such that $M \to^* N$. One says that the term $M$ \textit{has a normal form} if it reduces to a normal form in finitely many steps.

A reduction relation is \textit{confluent} when for every triplet $(M,N_1,N_2)$ of terms, the following implication holds:
\[M \to^* N_1 \wedge M \to^* N_2 \implies \exists M'.\, N_1 \to^* M' \wedge N_2 \to^* M'\]
Additionally, a reduction relation is said to be \textit{strongly normalizing} when every reduction sequence $M_0 \to M_1 \to \cdots$ eventually terminates.
\end{defi}

What follows is the operational semantics of the language FPC\@.

\[
\begin{prooftree}
\justifies
(\lambda x^\alpha.M)N \to M[N/x]
\end{prooftree}
\quad \quad
\begin{prooftree}
M \to M'
\justifies
\lambda x. M \to \lambda x. M'
\end{prooftree}
\quad \quad
\begin{prooftree}
M \to M',\, M \text{ not abstract}
\justifies
(M)N \to (M')N
\end{prooftree}
\]
(where an abstract term is a term of the form $\lambda x. M$ for some variable $x$ and some term $M$)
\medskip
\[
\begin{prooftree}
M \to N
\justifies
\op{inl}(M) \to \op{inl} (N)
\end{prooftree}
\quad \quad
\begin{prooftree}
M \to N
\justifies
\op{inr}(M) \to \op{inr} (N)
\end{prooftree}
\qquad
\begin{prooftree}
M \to \op{inl}(L)
\justifies
\opcase{M}{N}{P} \to N[x \mapsto L]
\end{prooftree}
\]
\medskip
\[
\begin{prooftree}
M \to \textbf{intro}_{\mu X. \tau}(N)
\justifies
\textbf{elim}(M) \to N
\end{prooftree}
\qquad
\begin{prooftree}
M \to \textbf{inr}(R)
\justifies
\opcase{M}{N}{P} \to P[y \mapsto R]
\end{prooftree}
\]

\end{document}